\documentclass[aps,prl,showpacs,twocolumn,groupedaddress]{revtex4}
\usepackage{graphicx}
\begin{document}

%\preprint{}

%\title{Entanglement and the Mott insulator - superfluid phase transition in linear chains with bosonic atoms.}
\title{Entanglement and the Mott insulator - superfluid phase transition in bosonic atom chains.}


\author{R. J. Costa Farias {\it and} M. C. de Oliveira}
%\thanks{}
\affiliation{ Instituto de F\'{i}sica ``Gleb Wataghin'',  Universidade
Estadual de Campinas, 13083-970, Campinas - SP, Brazil.}


\date{\today}

\begin{abstract}
We analyze the developing of bipartite and multipartite entanglement through the Mott-Insulator - Superfluid quantum phase transition. Starting from a Mott insulator state, where a filling factor $\nu = N/M = 1$ per lattice site is considered, we derive an exact expression for a completely connected graph configuration of bosons and show how bipartite and multipartite entanglement signals the phase transition predicted in previous works.  Moreover, through the transition bipartite entanglement shows to be monogamous.
\end{abstract}
%\keywords{}

\maketitle

%\section{}

Entanglement is a central resource of quantum mechanical systems and is particularly important for applications in quantum computation and quantum information science \cite{nielsen}. Recently it has been argued that entanglement may be quite relevant in many particle systems under quantum phase transitions (QPT) (See eg. \cite{amico08}). Indeed it has been proved that under certain conditions a non-analyticity appearing in a many-particle system ground state will be signaled in any bipartite \cite{sarandy04} and multipartite \cite{sarandy06,thiago06} entanglement measure. Several spin-1/2 models have been proved to follow this property. On the other hand, the interest in bosonic systems has increased recently due to the actual accessibility to several models in optical lattices experiments. One characteristic model that has attracted considerable attention is the Bose-Hubbard (BH) model under a Mott-insulator (MI) - Superfluid (SF) QPT. The concepts about phase transitions in the BH model were firstly derived by Fisher {\it et al.} \cite{Fisher1},  and experimentally achieved in a remarkable experiment by Greiner {\it et al.} \cite{Greiner}, becoming an intense object of investigation \cite{Jaksch2}. Some entanglement measures were explored previously in a wide variety of physical configurations \cite{Jaksch1,SougatoBose,Zurek,Buba}. At constant densities, an infinite order Berezinsky–-Kosterlitz--Thouless (BKT) quantum phase transition \cite{BKT} from the SF to the MI phase at low temperatures is expected for this system.  One important question arises: How is this infinite order QPT signaled by entanglement measures?

%In the present letter we develop this questioning by investigating a typical bosonic system suffering a  constant density MI- SF QPT. We investigate entanglement in two configurations deriving both analytical and numerical results: \textit{(i)} A  geometry referred as completely connected graph (CCG), where atomic interactions occur on-site and each atom is allowed to hop from one to any site of the whole lattice.  \textit{(ii)} A linear chain of bosonic atoms trapped in a 1D optical lattice. Atomic interactions occur on-site and each atom is allowed to hop to the two neighboring sites. This picture corresponds to the 1D BH model with mean number of bosons equal to 1. Both situations are analyzed considering a fixed filling number and consequently fixed density, here represented by the rate between the number of atoms ($N$) an sites ($M$), $\nu = N / M = 1$.
In this paper	we develop this questioning by investigating a typical bosonic system suffering a  constant density MI- SF QPT. We investigate entanglement in two configurations through analytical results: \textit{(i)} A  geometry referred as completely connected graph (CCG), where atomic interactions occur on-site and each atom is allowed to hop from one to any site of the whole lattice, Fig. (1)a.  \textit{(ii)} A linear chain of bosonic atoms trapped in a 1D optical lattice. Atomic interactions occur on-site and each atom is allowed to hop to the two neighboring sites, Fig. (1)b. This picture corresponds to the 1D BH model with mean number of bosons equal to 1. Both situations are analyzed considering a fixed filling number and consequently fixed density, here represented by the rate between the number of atoms ($N$) an sites ($M$), $\nu = N / M = 1$.
\begin{figure}
\begin{center}
\begin{tabular}{c c c}
%{\resizebox{\imsizeforthis}{!}{\includegraphics{new6comparacaode2.pdf}}}\\
%(a)\\
{   \includegraphics[height=2.5cm]{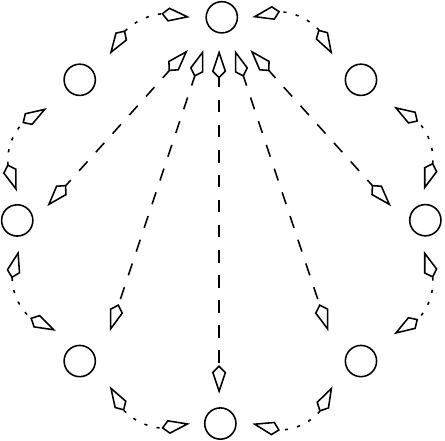}} & &{\includegraphics[height=2.5cm]{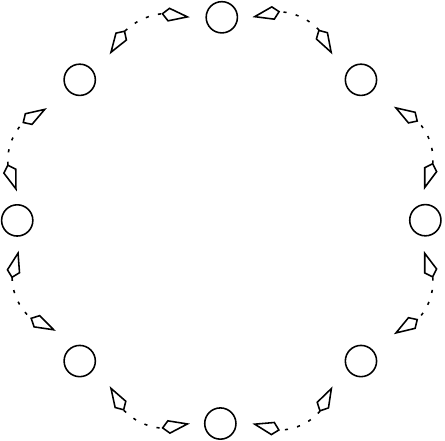}}\\
(a) &$\;\;\; $&(b)\\
%{\resizebox{\imsizeforthis}{!}{\includegraphics{6comparacaode4.pdf}}}\\
%(c)\\
\end{tabular}
\end{center}
\caption{ $N$- modes configuration. (a) CCG. (b) BH.}
\label{config}
\end{figure}

Let us consider the Hamiltonian for bosonic atoms in an external trapping potential
\begin{eqnarray}
\hat{H} &=&\int d^{3}r\hat{\psi} ^{\dagger }({\bf r})\left(-\frac{\hbar^2 }{2m}\nabla^{2} + V({\bf r})\right) \hat{\psi} ({\bf r})  \nonumber \\
&& + \frac{1}{2}U_0\int d^{3}r\hat{\psi} ^{\dagger }({\bf %
r})\hat{\psi} ^{\dagger }({\bf r})\hat{\psi} ({\bf r})\hat{\psi}
({\bf r}), \label{bec}
\end{eqnarray}
where $\hat{\psi}({\bf r}, t )$ and $\hat{\psi}^\dagger({\bf r},
t)$ represent the field operators, $V({\bf r})$ describes the optical lattice potential, $U_0 = 4 \pi\hbar^2a/m$ measures the strength of the two-body interaction, $m$ is the atomic mass and $a$ denotes the s-wave scattering length.
% Besides we use the reference \cite{VidalandWerner} to calculate the negativity for the present configurations.
The derivation of the BH or CCG Hamiltonian follows the standard procedures assuming all particles to be in the lowest band of the optical lattice \cite{Jaksch1,Jaksch2}. In this treatment, we expand the field operators in terms of the Wannier functions $\psi(x) = \sum_{i}\hat{a}_{i}w(x - x_{i})$, where $\hat{a}_{i}$ is the annihilation operator for a particle in site $x_{i}$ \cite{Jaksch1,Jaksch2}. Then, after some algebraic manipulation, we obtain the following Hamiltonian using an assumption that the total number of atoms $N$ is a conserved quantity
\begin{equation}
\mathcal{H} = U\sum_{i}N_{i}\left(N_{i} - I\right) - J\sum_{i,j}\left(\hat{a}^{\dagger}_{i}\hat{a}_{j} + \hat{a}^{\dagger}_{j}\hat{a}_{i}\right),
\label{bhm}
\end{equation}where $U = 4\pi a_{s}\hbar^{2}\int d^{3}x \left|w(x)\right|^{4}/m$ is the self-collision rate or the 
strength of the on site repulsion of two atoms on the lattice site $i$ and 
$J = \int d^{3}x w^{*}(x - x_{i})\left[-\left(\hbar^{2}\nabla^{2}\right) + V_{0}(x)\right]w(x - x_{j})$ is 
understood as the hopping matrix element between adjacent sites $i,j$ or the tunneling rate. The operators 
$\hat{N}_{i} = a^{\dagger}_{i}a_{i}$ count the number of bosonic atoms at the lattice site $i$; the annihilation and 
creation operators $a_{i}$ and $a^{\dagger}_{i}$ obey the canonical commutation relations $[a_{i},a_{j}^\dagger] = \delta_{i,j}$.
The BH Hamiltonian consists of (\ref{bhm}) taking the hopping term only between neighboring sites, while in the CCG configuration we consider hopping between any two sites. 
 Calculations through mean field theory, consistent with this last configuration, indicate the phase transition at the critical point of $U/J = 5.6z$, where $z = 2d$ is the number of nearest neighbors  
 \cite{Fisher1,Jaksch1,Jaksch2,Monien1}. For a 1D optical lattice, this mean field theory propose a ratio $J/U \approx 0.08$ to the transition point. These results were corroborated by the experimental investigation on phase transitions in an 1D optical lattice developed by St\"{o}rfele {\it et. al.} \cite{Stoferle}. Moreover there are well-known solutions for its two quantum phases \cite{Monien1}. Deep into the SF phase the system is described by a coherent state where the probability distribution for the local occupation of atoms on a single lattice site is Poissonian. Furthermore, this state is well described by a macroscopic wavefunction with long-range phase coherence throughout the lattice
\begin{equation}
\left|\Psi_{SF}\right\rangle = \frac{1}{\sqrt{N!}}\left(\frac{1}{\sqrt{M}}\sum_{i = 1}^{M}\hat{b_{i}}^{\dagger}\right)^{N} \left|0\right\rangle.
\label{sfsolution}
\end{equation}

In the MI phase the fluctuations in atom number of a Poisson distribution become energetically costly and the ground state of the system will instead consist of localized atomic wavefunctions with a fixed number of atoms per lattice site minimizing the interaction energy. The many-body ground state is then a product of local Fock states for each lattice site \cite{Jaksch1}
\begin{equation}
\left|\Psi_{MI}\right\rangle = \frac{1}{\sqrt{N!}}\prod_{i = 1}^{M} \left(\hat{b_{i}}^{\dagger}\right)^{n}\left|0\right\rangle
\label{misolution}
\end{equation}
Although there is a difference between the two configurations, BH and CCG, it is known that in the thermodynamic limit, 
  $N \rightarrow \infty$, the behavior of both is very similar. %That is, the CCG solution tends to the BHM one when $N \rightarrow \infty$.
 Experimentally it has been observed that the systems are sufficiently large with a number of sites $M$ of approximately $10^{5}$\cite{Greiner}. In this case the SF solution becomes indistinguishable from a coherent state that factorizes into a product of local coherent states of energy lattice site. This being a consequence of the commutation relation of the bosons at different sites \cite{Bloch}. 
The order parameter $\Delta N^2= <N^{2}> - <N>^{2}$ can be used as a signature of the transition. In fact, for MI (incompressible) phase $\Delta N^2=0$, showing us that the distribution is a Fock one, while in the SF (compressible) phase $\Delta N^2\neq0$, and a Poissonian distribution is observed.

The simplest way to investigate the evolution of entanglement through the QPT is to developed a perturbation treatment to determine the pure state of the system, departing from the MI ground state (\ref{misolution}) and taking the hopping term as a perturbation \cite{Monien1}. This so-called strong coupling expansion is valid since the transition is expected to occur at $J/U$ very small $ \approx 0.08$. Employing periodic boundary conditions and considering the unperturbed Hamiltonian as $\mathcal{H}_{0} = U\sum_{i}N_{i}\left(N_{i} - I\right)$ and $\mathcal{W} = \lambda\sum_{<i,j>} \left(\hat{a}^{\dagger}_{i}\hat{a}_{j} + \hat{a}_{j}^{\dagger}\hat{a}_{i}\right)$ where we define $\lambda = J /U$ as the perturbation parameter.
Since we are dealing with the microcanonical ensemble we neglect the chemical potential $\mu$ in our calculations. This is reinforced by the assumption that the total density of bosons is fixed. 
We perform our analytical calculations starting from a MI state considering an occupation number of exactly $\nu = N/M = 1$ atom per site.  We expect thus the MI-SF phase transition as soon as $\lambda$ departs from 0.

Since the system state is pure, multipartite entanglement (ME) can be detected by the reduced one-site linear entropy $S = \frac{d}{d - 1}\left(1 - Tr\rho_{N,1}^{2}\right)
$, where $d$ is the dimension of the relevant Hilbert space \cite{thiago06}.
Using the Equations (\ref{sfsolution}) and (\ref{misolution}) is possible to render an analytical expression for the reduced (one -site) estate and thus for the reduced linear entropy for the CCG configuration, considering a general normalized state of the system. For the CCG  it consists in distributing $N$ atoms in $M$ sites, considering the occupation number of each site as a label, and verify how many states are possible for the system: $|\psi\rangle\equiv|\mathcal{D}\rangle= \sum_i a_i |i\rangle$, where $\mathcal{D}$ is the combinatorial form to distinctly distribute $N$ particles in $M=N$ sites, 
%\begin{eqnarray}
%\mathcal{D} = \left(\begin{array}{c} N + M - 1 \\ N \\ \end{array} \right),
%\label{algeff1}
%\end{eqnarray}
%which in the present case $N = M$, corresponds to
\begin{eqnarray}
\mathcal{D} = \left(\begin{array}{c} 2N - 1 \\ N \\ \end{array} \right).
\label{algeff2}
\end{eqnarray}
 Here $a_i$ is a complex parameter characterizing the contribution  of each combinatorial state $|i\rangle$ in the description of the physical state $|\psi\rangle$. $a_i$ is related to the correlations of a site with the rest of the lattice. Since in this model we deal with only two-body interaction $a_i$ is indeed related to two-body correlations, thus bringing all the information about the reduced system. Therefore it certainly is a function of the imbalance between the interaction term $U$ and the hoping term $J$. Whether it is continuous or not is what determines the type of the phase transition from MI to SF. In our approach, $a_i$ is obviously a truncated power series over the perturbation parameter $\lambda$. The least costing global state starts in the MI phase with one single atom occupying each site: $|111\cdots1\rangle$, whose parameter is $a_1$. After some combinatorial reasoning and partial trace over $N-1$ sites the general relation for the reduced one-site state for $N = M$ is given by
\begin{equation}
\rho_{N,1} = \left(N-1\right)!\sum_{i = 1}^{f(N)}\sum_{j = 0}^{N}\left|\alpha_{i}\right|^{2}A_{ij}\left|j\right\rangle\left\langle j\right|,
\label{finalrho}
\end{equation}
where $f(N)$ is the integer partition function of the number $N$, and %\begin{equation}
$A_{ij} = r_{i}^{j}/\prod_{\ell}r_{i}^{\ell}$ %,\end{equation}
 represents all the possibles arrangements for the symbols $r_{i}^{j}$, necessary to write down the occupancy  $j$ of a particular site. Eq. (\ref{finalrho}) satisfies the constraint determined by $\left|\alpha_{i}\right|^2 \geq \left|\alpha_{i + 1}\right|^2$, that is understood as a consequence of the energetic cost since the on-site interaction becomes very strong and leads to a highly unstable state as the occupation in a determined site is increased.
Now the reduced one-site linear entropy, $S$, can be calculated from Eq. (\ref{finalrho}) and plotted in function $\lambda$ and $N$ to up to the required order, as in Fig. (2).

Before we move on to discuss the entropy, let us make some remarks. Firstly, we notice that in principle, with increasing $N$ it will be required higher orders of perturbation to describe the exact state. However from the hierarchy of $a_i$ above those terms are likely to be very small, and thus a first or second order perturbation is  invariably enough for the description of the behavior of ME. Secondly we remark on the possibility of $S$ signaling the QPT. Analyzing the first derivative in $\lambda$ of the $S$ we find  in the thermodynamic limit $N \rightarrow \infty$,
\begin{equation}
\frac{\partial S}{\partial \lambda} = - 2 \frac{d}{d - 1} \sum_{i = 1}^{\infty} \left|\alpha_{i}\right|^{2} \sum_{j = 0}^{\infty}A_{ij} \frac{\partial\left|\alpha_{i}\right|^2}{\partial \lambda}.
\label{nonanaliticities}
\end{equation}
In this limit, one has to consider an infinite occupation number and thus an infinite number of $a_i$'s each one of the $|\alpha_{i}|$ tends to zero and in this same limit $A_{ij}$ is just a constant. Thus each $i$-th term in the summation is close to zero unless the derivative in $a_i$ is divergent or at least very large.  For a continuous QPT it is expected that $\partial|\alpha_{i}|^{2} / \partial\lambda$ be the origin of a discontinuity or divergence if the transition is of the second order \cite{thiago06}. For an infinite order QPT although the derivative in $a_i$ may be large it is finite, but still can signal the QPT. The same reasoning is valid for derivatives of higher orders, which are then continuous. On the other hand the energy corresponding to the system perturbed state $E(\lambda)=\sum_n \lambda^n {\cal E}^{(n)}$, where ${\cal E}^{(n)}$ is the $n$-order energy, is always continuous on $\lambda$ as well, but it is unlikely to contain any non-analyticity, and thus will not signal the QPT.

In Fig (\ref{fig345678}) the linear entropy and its derivative are plotted  for a number of atoms up to 1000 for both the CCG and the BH configurations, employing a perturbation up to second order. To start up, we exemplify in Figs. (\ref{fig345678})a and b the two simplest situations, when $N = M = 2$ and $3$, where the BH and the CCG configuration are indistinguishable.
\newcommand{\imsizeforthis}{0.47\columnwidth}
\begin{figure}
\begin{center}
\begin{tabular}{c c}
{\resizebox{\imsizeforthis}{!}{\includegraphics{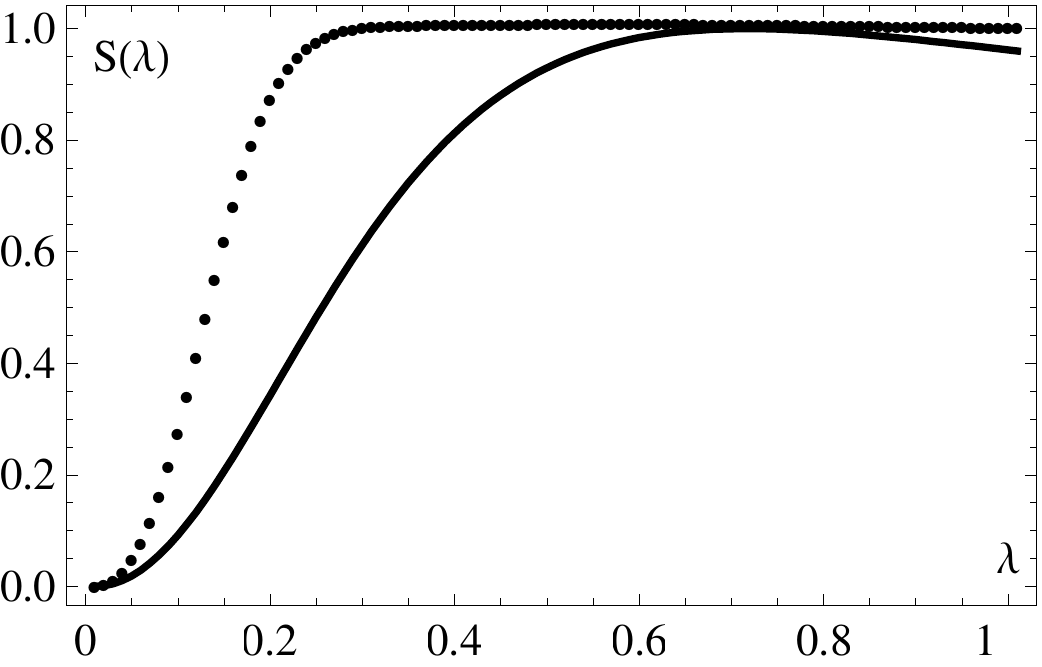}}} & {\resizebox{\imsizeforthis}{!}{\includegraphics{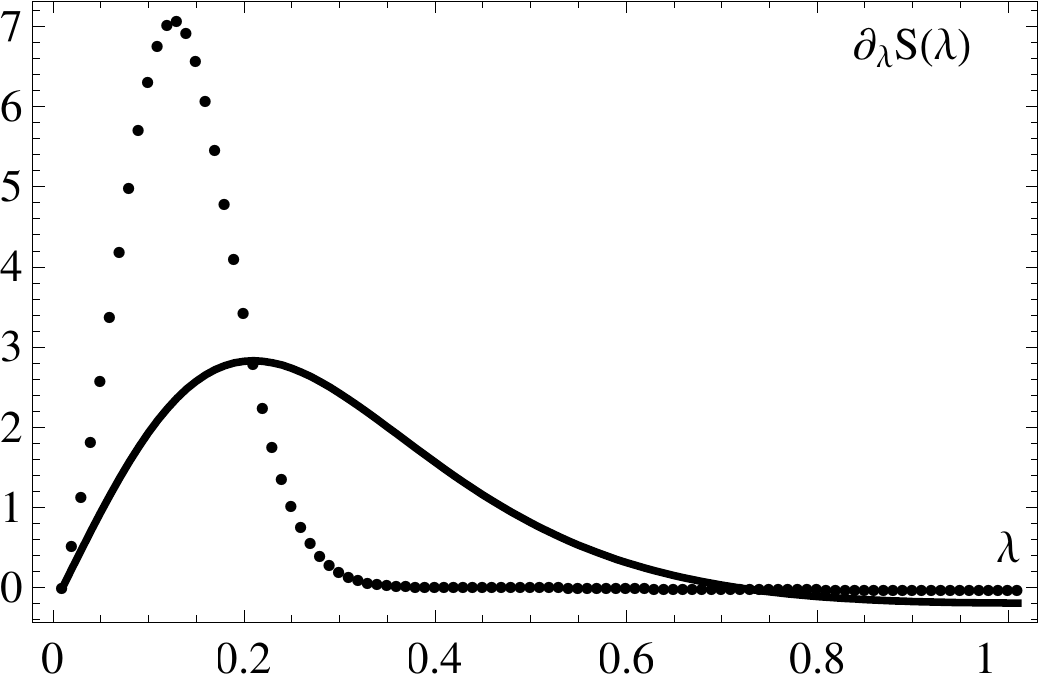}}}\\
%{\resizebox{\imsizeforthis}{!}{\includegraphics{Comparison2and3.pdf}}}\\
(a) & (b)\\
%{\resizebox{\imsizeforthis}{!}{\includegraphics{Compdev23.pdf}}}\\
%(b)\\
{\resizebox{\imsizeforthis}{!}{\includegraphics{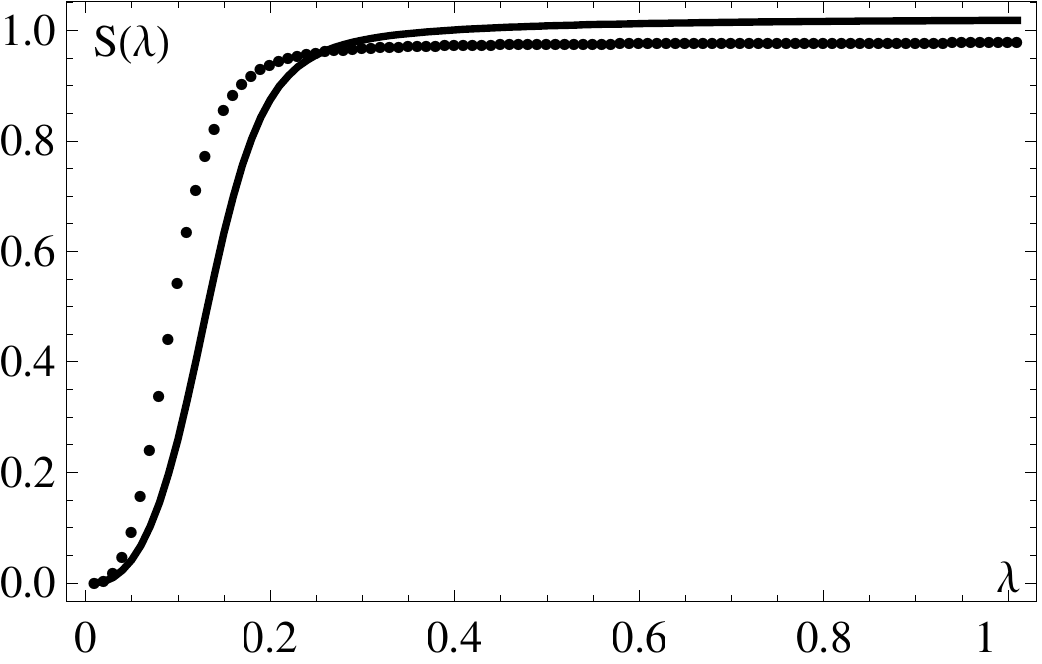}}} & {\resizebox{\imsizeforthis}{!}{\includegraphics{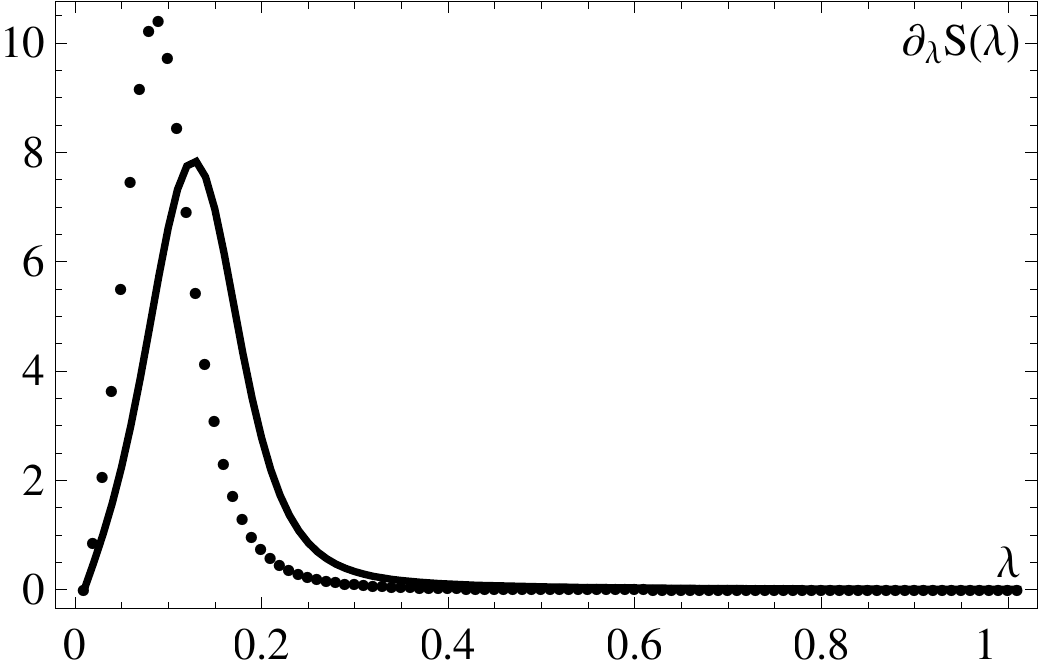}}}\\
%{\resizebox{\imsizeforthis}{!}{\includegraphics{Comp4GCCBHM.pdf}}}\\
(c) & (d)\\
%{\resizebox{\imsizeforthis}{!}{\includegraphics{Compdev4.pdf}}}\\
%(d)\\
{\resizebox{\imsizeforthis}{!}{\includegraphics{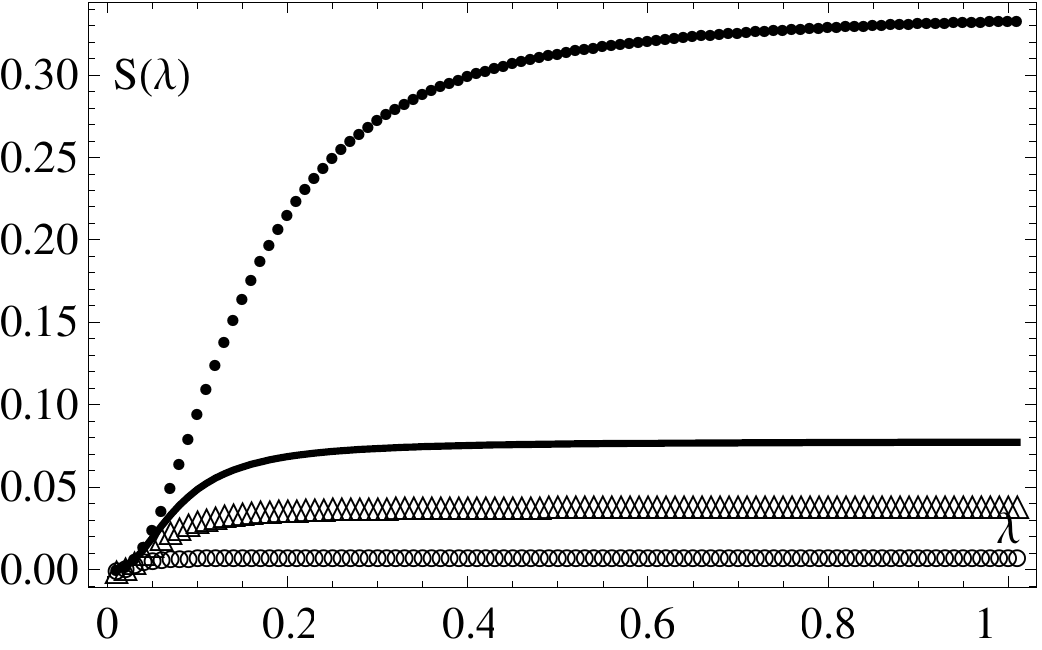}}} & {\resizebox{\imsizeforthis}{!}{\includegraphics{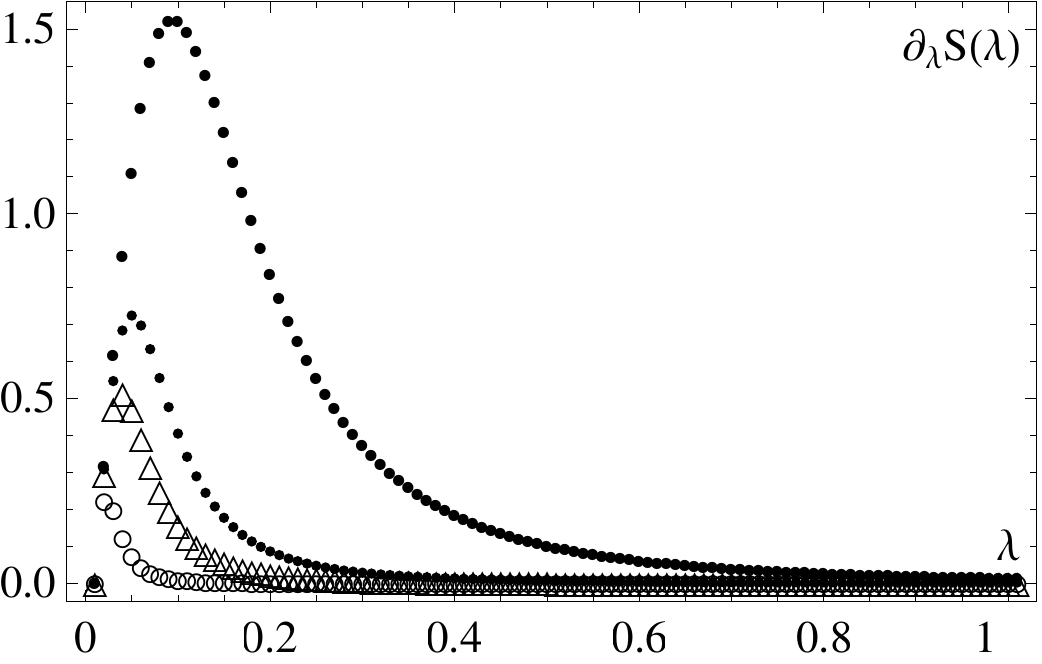}}}\\
%{\resizebox{\imsizeforthis}{!}{\includegraphics{uniaofigBHM.pdf}}}\\
(e) & (f)\\
\end{tabular}
\end{center}
\caption{(a) Reduced one site linear entropy, and (b) correspondent derivatives, as a function of $\lambda$ for $N = M = 2$ (solid line) and $3$ (dotted line) for the 1D-BH and the GCC configuration. (c) Reduced one site linear entropy, and (d) respective derivative, as a function of $\lambda$  for $N = M = 4$ situation in 1D-BH (solid line) and CCG (dotted line) configuration.(e) Reduced one site linear entropy, and (f) corresponding derivative, as a function of $\lambda$ for $N = M =$ 10 (dotted line), 50 (solid), 100 (triangles) and 1000 (circles) calculated up to the first order for the 1D-BH configuration.}
\label{fig345678}
\end{figure}
For $\lambda$ small the system can be represented as a product of local states, with no entanglement. However, as $\lambda$ is increased, we observe an increase of the linear entropy, stabilizing after $\lambda\approx 0.3$ for $N=3$. This means that as the system tends to the SF phase, its state no longer is separable, and consequently the entanglement is increased. 
%The first situation where the evolution of the two configurations reduced one site linear entropy are different occurs when $N = M = 4$. In this cases, the states calculated show a distinct evolution because of the different form of the perturbation part of the Hamiltonian, where this time the hopping can occur between any two sites for the GCC configuration. The result for these cases are showed in Figures (\ref{2and3a}b) and (\ref{2and3a}c). In Figures (\ref{2and3a}e) and (\ref{2and3a}f) we exemplify as we can get the same result for a large number of bosons and sites for the BHM. Although is not possible to determine the critical point through our calculations, because it is feasible just in a grand canonical ensemble, we will show that it is qualitatively perfect to observe the transition through the reduced one site linear entropy and of the negativity as a function of the perturbation parameter.
Different behaviors for the BH and CCG configurations occur when $N = M = 4$ and so on, as can be viewed in Figs. (\ref{fig345678})c and d. These distinct evolutions are consequence of the different form of the perturbation term of the Hamiltonian.  In Figs. (\ref{fig345678})e and (\ref{fig345678})f we show for the BH model employing first order perturbation only, for several distinct number of bosons.
%Although is not possible to determine the transition point through our calculations, because it is feasible just in a %more appropriated method, as can be an approach in a grand canonical ensemble, it is qualitatively perfect to observe %the transition through the reduced one site linear entropy and of the negativity as a function of the perturbation %parameter.
%It is interesting to note that the
 The linear entropy characterizes very well the two phases predicted by the BH model. 
When $\lambda=0$ we observe no entanglement, characterizing the MI phase. In the other limit, as $\lambda\rightarrow 1$ the state tends to be maximally entangled. This is a typical behavior of a system in the SF phase. We remark that the linear entropy profile is very similar to the order parameter $\Delta N^2$ (not shown). Thus $\Delta N^2$ is indeed a witness of the ME.
Although we cannot do strong assumptions about the point where the phase transition occurs, given the perturbative approach, the behavior of ME  signals the QPT correctly.

%\section*{{\em Negativity}}

\begin{figure}
\begin{center}
\begin{tabular}{c c}
%{\resizebox{\imsizeforthis}{!}{\includegraphics{new6comparacaode2.pdf}}}\\
%(a)\\
{\resizebox{\imsizeforthis}{!}{\includegraphics{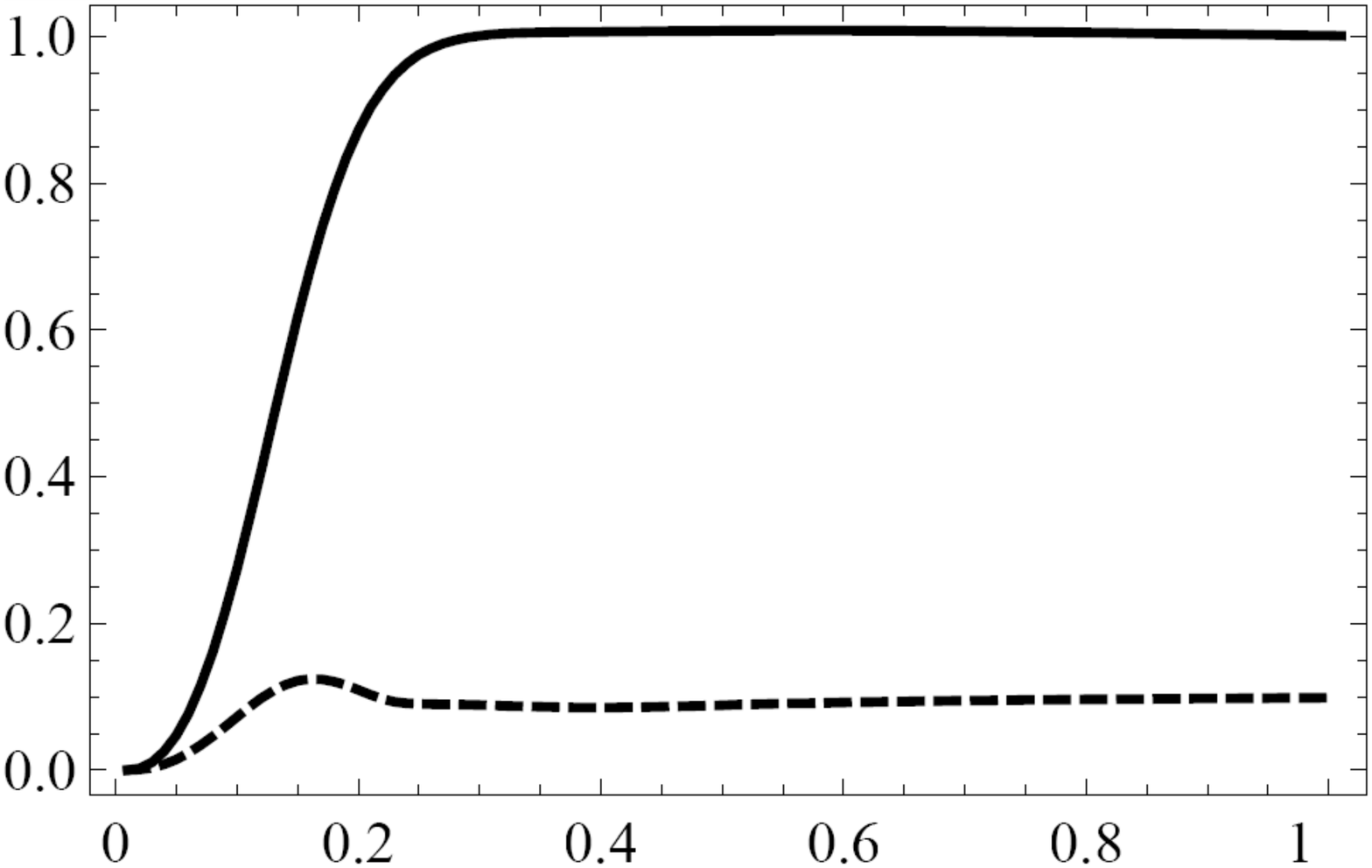}}} & %{\resizebox{\imsizeforthis}{!}{\includegraphics{6comparacaode4.pdf}}}\\
{\resizebox{\imsizeforthis}{!}{\includegraphics{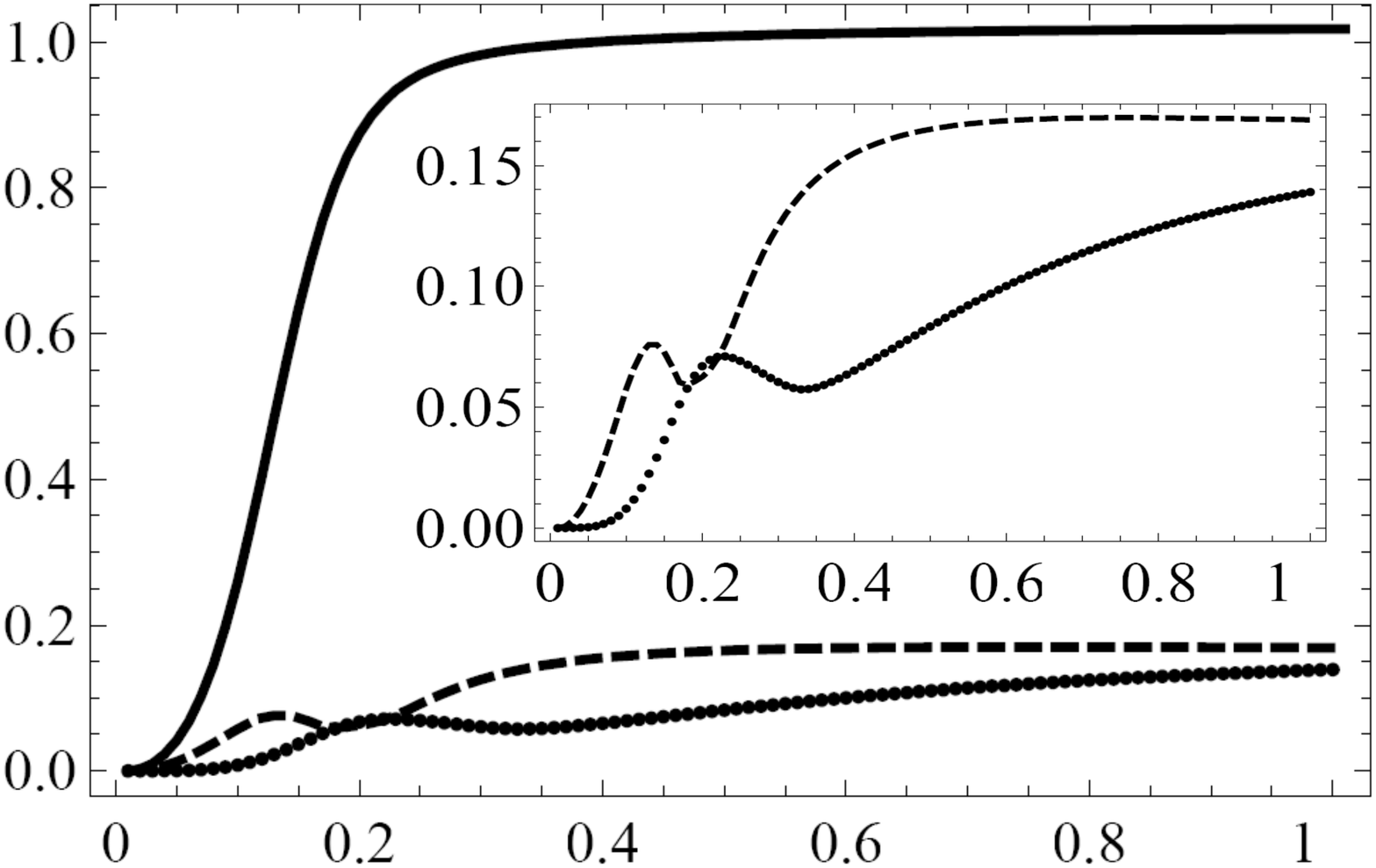}}}\\
(a) & (b)\\
%{\resizebox{\imsizeforthis}{!}{\includegraphics{6comparacaode4.pdf}}}\\
%(c)\\
\end{tabular}
\end{center}
\caption{ (a) $\mathcal{N}_{12}$ (dashed line) for $N = M = 3$ and the reduced one site linear entropy (solid line) and (b)$\mathcal{N}_{12}$ (dashed line), $\mathcal{N}_{13}$ (dotted line) and the reduced one site linear entropy  for $N = M = 4$.}
\label{fig910}
\end{figure}

We quantify bipartite entanglement (BE) through the negativity \cite{VidalandWerner}, defined as $\mathcal{N}(\rho) \equiv \frac{\left\|\rho^{T_{A}}\right\|- 1}{2}$, where $\left\|\rho^{T_{A}}\right\|$ is the trace norm of the partially transposed state $\rho_{i,j}^{T_{A}}$ of any pair of sites  $\{i,j\}$ for the BH model. This is equivalent  to the absolute value of the sum of negative eigenvalues of $\rho^{T_{A}}$, vanishing for separable states. Although strictly necessary and sufficient only for Hilbert space dimension up to $2\otimes 3$, in the present case it is correctly characterizing the BE for any $N$.
In Fig. (
\ref{fig910}) we plot the negativity and the linear entropy for $N=3$ and $N=4$. Since for $N=3$ there are only nearest neighbor for any site the negativity is the same for any two sites. For $N = M = 4$ the negativity shows different behavior for nearest and next-nearest neighbors, here exemplified by $\rho_{12}$ and $\rho_{13}$, respectively. Contrary to spin-1/2 model (see e.g.\cite{Oliveira3}) the next-nearest neighbor negativity cannot be neglected in comparison to the nearest neighbors one. Instead they show a peculiar behavior signaling the monogamy of entanglement.
%As we can see in the Figure (\ref{negativityfigure}b) we observe a decreasing in the curve corresponding to the $\rho_{12}$ at the same interval that $\rho_{13}$ increases as a function of the perturbation parameter. We interprete this behavior as a consequence of an inequality involving these quantities because of the differents modalities of enetanglement, showing us that this behavior is a signature of the monogamy of entanglement to the BHM.
In the simplest situation ($N = M = 2$), since the system is pure, the negativity is equal to the one site reduced linear entropy as there is only BE. When $N = M = 3$ the negativity after showing an increase in the transition, stabilizes at lower values than the one site linear entropy. This is a signature that genuine tripartite entanglement do exist in the system. Indeed the ME develops through the bipartite one (the two curves evolve similarly for small $\lambda$). But as soon as tripartite entanglement starts to develop the two curves diverge. 
In the $N = M = 4$ case, the negativity for $\rho_{12}$ and $\rho_{13}$ shows that distinct kind of BE exists. ${\cal N}_{12}$, after increasing with the entropy, decreases as ${\cal N}_{13}$ increases (see the inset in Fig. 2), showing an interesting bound on the distribution of BE. Subsequently  both measures stabilize to closer values. This bound on BE is typical of the monogamy of entanglement \cite{Wooters} in the sense that the increasing  of the $\rho_{12}$ entanglement at the same interval where the $\rho_{13}$ one decreases and a reciprocity maintain a constant amount of the BE. Here the difference between the ME and the two types of BE, namely the residual entanglement signals both tripartite and quadripartite entanglement. This feature continues for increasing $N$, and there is an increasing number of types of BE. The initial crossing of BE decreases with N, stabilizing after a while at higher values. This is an evident manifestation of the many classes of entanglement present in the BH model. The SF phase is intermediated by a strong ME state where any mode (site) is entangled with the others in many distinct ways.

In short, we analyzed ME and BE for the CCG and  1D-BH configuration of bosonic atoms trapped in an optical lattice. We described the behavior of entanglement through the MI-SF QPT predicted for these models showing how it is signalized by two entanglement measures, the reduced one site linear entropy and negativity.
By employing the linear entropy just the diagonal elements of the whole density matrix must be taken into account, since the partial trace keeps only these terms (see Eq. (\ref{finalrho})). It represents a remarkable reduction of numerical resources for investigation of a QPT in comparison with the full Hamiltonian diagonalization.
 For instance, for $N = M =$ 10, by Eq.  (\ref{algeff2}) one must diagonalize a  92378 $\times$ 92378 matrix, while through the linear entropy it involves the calculation of just 42 coefficients in the case of GCC configuration, which  can be solved by some computational method with relative simplicity.

The authors would like to acknowledge A. Schwartz and E. Miranda for enlightening discussions. This work is supported by CNPq and FAPESP.

\end{document}